# Crossover between kite growth and vibrational bridging in pillar-assisted controlled formation of carbon nanotube networks

*Yuanjia Liu [a], Taiki Inoue [a], Yoshihiro Kobayashi [*a]*

a Department of Applied Physics, The University of Osaka,

Suita, Osaka 565-0871, Japan

*Email: kobayashi@ap.eng.osaka-u.ac.jp

Abstract

   Pillar-assisted growth is a technique in which short carbon nanotubes (CNTs) form suspended networks by growing across closely spaced microfabricated pillars. During growth, the CNT tips exhibit vibrations that allow them to bridge the neighboring pillars. To improve the complexity and controllability of the CNT networks, we introduce a kite-growth mechanism in which CNTs are elongated and aligned by gas flow during growth, enabling a longer bridging distance compared to vibrational bridging. By integrating theoretical modeling, simulations, and experimental synthesis, we found that CNT tip vibrations dominate bridging at short lengths, whereas gas flow increasingly influences alignment as CNTs grow longer. This results in a crossover behavior governed by gas flow and pillar arrangement. We also developed a bridging model based on geometric constraints to quantify the bridging behavior based on pillar spacing and angular accessibility. The statistical analysis of the resulting network structures demonstrates that the pillar arrangement significantly influences the connection types, with kite growth enabling more diverse network topologies. These findings provide design principles for tuning the density and structural complexity of suspended CNT networks, offering promising applications in nanoscale electrical interconnect wiring and three-dimensional circuit architectures.

Keywords: carbon nanotubes, networks, nanopillar, kite growth, chemical vapor deposition, finite element method

# 1. Introduction

Carbon nanotubes (CNTs) [1,2] possess exceptional mechanical strength, electrical conductivity, and thermal stability at the individual level [3–5]. However, using them as isolated elements poses challenges in practical applications. To overcome this issue, CNTs are often assembled into interconnected networks, which not only preserve their intrinsic properties but also enhance their flexibility, durability, and large-area scalability. These CNT networks have attracted significant attention for applications in nanodevices [6], energy storage [7], and sensing technologies [8]. In particular, both aligned arrays and random CNT networks have been actively investigated as potential interconnect wiring or electrodes in next-generation electronics [9–12]. Although techniques, such as spin coating and inkjet printing [13,14], have been used to fabricate CNT networks, they often offer limited control over network density and bridging configurations, which are crucial for consistent electrical characteristics in large-scale integration.

Pillar-assisted growth provides spatial anchors for CNTs, enabling the formation of suspended structures and geometric control over their orientation and connectivity. Originally proposed by Dai et al. [15,16], this method adopts regularly spaced nanopillars during chemical vapor deposition (CVD) to guide the self-assembly of CNTs into suspended bridges. Dai's work used microscale pillar spacings (typically ~5–10 μm) and achieved remarkably long CNT spans (up to 150 μm), attributing their formation primarily to the directionality of the gas flow. Notably, although the concept of kite growth [17–21] had not yet been formalized at the time, many observed features in Dai's work strongly reflect its characteristics, such as ultralong CNTs with strong alignment along the gas flow direction and bridging across multiple pillars. This mechanism, characterized by tip growth influenced by buoyant gas flow forces, allows CNTs to extend into higher laminar flow zones and form longer suspended structures. In contrast, Homma et al. later proposed a vibrational growth model in which CNTs oscillate during growth and connect to nearby pillars. Their experiments, using submicron spacing pillar arrays (400–1,400 nm) [22–25], demonstrated that CNT tip vibrations were sufficient to account for short-range bridging, challenging the earlier gas-flow-based explanation [16]. Both interpretations are valid within their respective experimental regimes—Dai under conditions favoring extended, flow-guided growth [16], and Homma under conditions where vibrational growth likely dominates the growth behavior [22]. More recently, Lee et al. expanded the scope of pillar-assisted growth to three-dimensional CNT networks with tunable density and connectivity [26–29]. Their work highlighted the potential of nanopillar arrays as flexible platforms for customized interconnect architectures, with applications in multi-I/O routing and device wiring [30,31]. Although their study did not explicitly analyze the underlying growth mechanism, their experimental conditions suggest a regime in which vibrational growth dominates. Overall, previous studies have primarily focused on regimes where either flow-guided or vibrational growth dominates, without investigating the transitional or overlapping regimes where both mechanisms coexist or compete. Understanding these crossover conditions is essential for the precise control of the CNT bridging behavior.

In this study, we developed a unified growth framework that explicitly incorporates kite growth into a pillar-assisted technique for fabricating structurally tunable CNT networks. Building on our previous demonstration of kite growth [21], we systematically investigated the influence of the gas flow direction and pillar spacing on the CNT bridging behavior using extensive experimental data and fluid dynamics simulations. Our analysis provides insight into the interplay between the vibrational and kite-growth mechanisms and demonstrates how a controlled-array design can optimize network connectivity. The resulting controllability of the CNT bridging structures may also facilitate broader applications, including single-tube characterizations [32] and scalable nanoscale wiring [30].

## 2. Experimental section

### 2.1 CNT growth and characterization

Quartz substrates were patterned to form $SiO_2$ nanopillar arrays via a microfabrication process. The nanopillars had a square cross-section with a side length of 1.6 μm and a height of 8 μm. The edge-to-edge spacing between the nearest neighbor (NN) pillars was set to 3, 5, and 10 μm to assess the influence of the pillar arrangement on the CNT network formation.

A 0.03 M ethanol solution of $FeCl_3$ was spin-coated onto the substrates to serve as a catalyst precursor. After catalyst deposition, the substrates were placed in a horizontal CVD furnace for thermal treatment and CNT growth. Under a continuous flow of 20 sccm $H_2$ (3%)/Ar at 85 kPa, the temperature was ramped to 850°C–950°C and held for 10 min to reduce the catalyst. Subsequently, ethanol vapor was introduced as the carbon source. The ethanol vapor was generated by bubbling 35 sccm $H_2$ (3%)/Ar through a 40°C ethanol reservoir, supplemented by an additional 420 sccm $H_2$ (3%)/Ar flow through a separate line. CNT growth proceeded for 30 min under atmospheric pressure, with an estimated partial ethanol pressure of approximately 1.4 kPa. The gas flow was horizontally directed across the substrate surface throughout the process. The growth conditions were designed to promote tip-growth behavior consistent with previously established kite-growth mechanisms [21], thereby facilitating the formation of suspended CNT networks that bridge the nanopillar arrays. After growth, the reaction was terminated by switching the gas flow to pure Ar and cooling the furnace to room temperature under an Ar atmosphere.

The morphology and structural features of the CNT networks were characterized using scanning electron microscopy (SEM). SEM images were obtained using a Hitachi S-4800 system with an accelerating voltage of 5 kV and an emission current of 10 μA. SEM images were acquired at tilted angles to provide three-dimensional views of the CNT morphology, with the sample rotated horizontally by 30° and tilted vertically by 30°.

### 2.2 Finite element method simulations

Finite element method (FEM) simulations were conducted using COMSOL Multiphysics (COMSOL Inc.) to investigate the gas flow behavior and mechanical response of CNTs during growth.

The Laminar Flow module was used to model the gas flow field around the nanopillar arrays. The simulation solved the steady-state incompressible Navier–Stokes

equations as follows:

$$\nabla \boldsymbol{u} = 0, \quad (1)$$

$$\rho(\boldsymbol{u} \cdot \nabla)\boldsymbol{u} = \nabla \cdot \sigma + \boldsymbol{F}, \quad (2)$$

where $\boldsymbol{u}$ denotes the velocity field, $\rho$ denotes the fluid density, and $\boldsymbol{F}$ represents the body force per unit volume. The Cauchy stress tensor $\sigma$ is defined as

$$\sigma = -p\boldsymbol{I} + \mu[\nabla \boldsymbol{u} + (\nabla \boldsymbol{u})^T], \quad (3)$$

where $p$ denotes the pressure, $\mu$ denotes the dynamic viscosity, and $\boldsymbol{I}$ denotes the identity tensor. The boundary conditions were set to match the experimental gas flow velocity (0.01 m/s) with no-slip conditions applied at the nanopillar surfaces. Mesh refinement was applied near the nanopillars to accurately capture the local velocity gradients. Given the low flow velocity and microscale geometry, the flow was assumed to be laminar and incompressible throughout. Further details on the simulation parameters, mesh configurations, and material properties are provided in the Supporting Information.

## 3. Results and discussion

### 3.1 Gas flow direction dependence and model fitting

CNT networks were fabricated on substrates consisting of nanopillar arrays with various geometric layouts under different growth conditions. For simplicity, the directions that are parallel and perpendicular to the gas flow direction are designated as // and ⊥, respectively. These symbols specify the orientation of the nanopillar configuration relative to the gas flow direction. For consistency, samples are designated based on the spacing of nanopillars in the parallel (//) and perpendicular (⊥) directions to the flow; e.g., a sample labeled 3 × 10 corresponds to nanopillar spacings of 3 μm (//) and 10 μm (⊥).

SEM analysis (Fig. S2) revealed that the catalyst particles were likely distributed across the entire pillar surface due to the liquid coating process. However, CNTs that nucleate at the apexes can effectively contribute to suspended bridging. CNTs that grow from the sidewalls of pillars or the substrate tend to quickly adhere to nearby surfaces, which prevents them from contributing to network formation. Using anisotropic pillar configurations and past research, the concept of "nearest neighbors between pillars" is modified and refined to characterize bridging patterns in CNT networks [22]. The parallel NNs (//NN) and perpendicular NNs (⊥NN) are nanopillars that are exactly near each other in either the parallel (//) or perpendicular (⊥) directions. The diagonally aligned pillars are classified as diagonal NNs (∠NN). A schematic illustration of these configurations and bridging modes is shown in Fig. 1. Since this study focused on suspended CNT structures, CNTs that adhered to the substrate surface were excluded from the analysis.

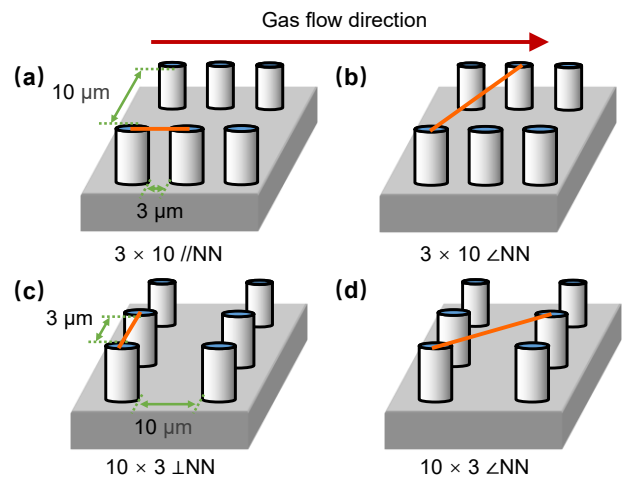

**Figure 1 Schematic illustration of the CNT bridging behavior in different nanopillar configurations.**

**(a, b) and (c, d) correspond to pillar arrays with short spacing aligned parallel (3 × 10) and perpendicular (10 × 3) to the gas flow, respectively. (a) and (c) represent //NN and ⊥NN connections, whereas (b) and (d) represent ∠NN. The red arrow indicates the gas flow direction.**

Figs. 2(a) and (b) show the representative SEM images of the CNT networks synthesized under kite growth conditions [21] using substrates with identical nanopillar arrangements but subjected to distinct gas flow directions, explicitly comparing the 3 × 10 and 10 × 3 configurations. In both circumstances, the CNTs mostly bridged across the adjacent nanopillars; however, depending on the direction of the gas flow, minor variations in the bridging behavior were observed. Although the overall network structures appeared similar in the SEM images, subtle differences in the bridging behavior were further analyzed quantitatively. Specifically, two groups of samples, including those shown in Figs. 2(a) and (b), were systematically evaluated. The // group comprises the 3 × 10, 5 × 10, and 10 × 10 configurations in which the nanopillar spacing in the // direction is altered (3, 5, and 10 μm) while preserving a constant 10 μm spacing in the ⊥ direction. The ⊥ group includes the 10 × 3, 10 × 5, and 10 × 10 samples in which the ⊥ direction spacing varies, while the // spacing remains constant at 10 μm. Fig. S3 shows typical SEM images of the CNT networks formed in the 5 × 10, 10 × 5, and 10 × 10 layouts. Fig. 2(c) summarizes the bridging statistics as a function of pillar spacing for both sample sets in the respective gas flow directions. These statistics represent the average number of CNT bridges per nanopillar pair, classified by connection type—specifically, //NN, ⊥NN, and ∠NN categories defined in the previous paragraph. As shown in Fig. 2(c), two main trends are evident. First, as the spacing between the pillars increased, the likelihood of CNT bridging decreased, leading to a reduced number of bridges in both groups. Second, the direction of the gas flow affects the //NN and ⊥NN bridging connections, whereas its impact on ∠NN connections appears minimal. For each group, the number of CNT bridges in the // direction was consistently higher than in the ⊥ direction.

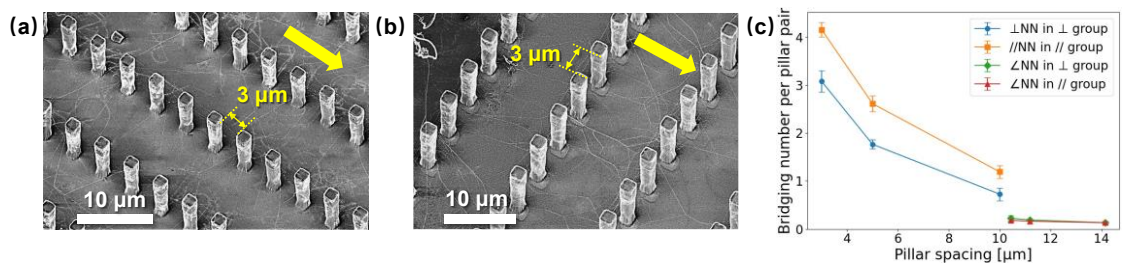

**Figure 2 (a, b) SEM images of the CNT networks grown on substrates with identical nanopillar**

**arrangements but different gas flow directions. (a) Sample with the 3 × 10 configuration, that is 3 μm spacing in the // direction and 10 μm spacing in the ⊥ direction. (b) Sample with the 10 × 3 configuration, that is 10 μm spacing in the // direction and 3 μm spacing in the ⊥ direction. The yellow arrows indicate the direction of the gas flow during growth. (c) Pillar-spacing dependence of the average number of CNT bridges per pillar pair, categorized by //NN, ⊥NN, and ∠NN connections in both // and ⊥ groups. For each pillar spacing, we grew multiple samples, captured SEM images at several regions per sample, counted bridges per pillar pair in each image, averaged these values, and plotted the standard deviation as the error bar.**

The observed anisotropy in the bridging behavior is attributed to the interplay of two main growth mechanisms: vibrational growth, which enables both //NN and ⊥NN bridging [23], and kite growth, which predominantly promotes //NN along the gas flow direction [21]. The environmental SEM has directly detected the vibratory action in situ, indicating the validity of the proposed vibrational growth model [24]. In contrast, the kite growth process primarily occurs under directional gas flow conditions. Here, CNTs are lifted from the tops of the nanopillars and carried along by laminar flow, which enables them to bridge greater distances and span multiple nanopillars [21].

To quantitatively describe the CNT bridging behavior, we revisited a previously proposed isotropic bridging model [23], which assumes that CNTs grow straight without any directional preference in any direction from the pillar top and contact an adjacent pillar. Fig. 3(a) illustrates the geometric definitions of the two angular constraints involved in CNT bridging between nanopillars: the transverse angle $\Omega$, representing the horizontal angular coverage from one pillar top toward a neighboring one, defined by a tangential projection from the pillar edge, and the longitudinal angle $\Omega'$, characterizing the vertical accessibility range from the pillar top to a neighboring pillar defined from the tangential point down to the base for successful bridging. In the original model, the bridging probability $P$ is assumed to scale with the transverse angle as [23]:

$$P = K\Omega^n, \tag{4}$$

where $K$ represents a proportionality constant, and $n$ represents a fitting exponent. It is worth noting that the "bridging probability" used in previous studies was based on binarized counts (0 or 1) by limiting the catalyst density so that only a single CNT bridge could form per pillar pair. In contrast, this study does not impose such constraints; thus, multiple CNTs may bridge a single pillar pair, leading to values greater than 1. For a consistent comparison across different experimental approaches, we define $B$, the average bridging number per pillar pair, as the primary metric. Unlike the probability values $P$ used in previous studies that limited CNT counts per pair, $B$ generalizes the bridging behavior to reflect the actual CNT density, including multibridge events.

Based on the geometric constraints, we derive an analytical expression for the bridging number per pillar pair $B$ as

$$B = k\Omega \cdot k'\Omega' = K\Omega\Omega', \tag{5}$$

where $k$ and $k'$ denote conversion factors translating the available transverse and longitudinal angles into bridging numbers, and $K$ represent their combined coefficients. The angles $\Omega$ and $\Omega'$ are explicitly given by

$$\Omega = 2\arcsin\left(\frac{d}{2(L+d)}\right), \tag{6}$$

$$\Omega' = \arctan\left(\frac{2h\tan(\Omega/2)}{d\cos(\Omega/2)}\right), \tag{7}$$

where $L$, $d$, and $h$ denote the interpillar spacing, pillar diameter, and height, respectively. The detailed derivation is provided in the Supplementary Information Section S1. By substituting Eqs. (6) and (7) into Eq. (5) and using the corresponding trigonometric relations, we derive the following analytical form:

$$B = 2K \cdot \arcsin\left(\frac{d}{2(L+d)}\right) \cdot \arctan\left(\frac{h(L+d)}{(L+d)^2 - (d/2)^2}\right). \tag{8}$$

This geometric formulation provides a systematic framework for evaluating the bridging number based on and expanding previous simplified approaches. Given fixed pillar dimensions ($d$ and $h$), the trend of $B$ as a function of spacing $L$ is uniquely determined by the model. The parameter $K$ summarizes the effective density of the CNTs. A higher CNT density during growth increases the successful bridging number, thereby increasing the value of $K$. Thus, this formulation allows $B$ to serve as a geometry-dependent indicator of bridging behavior.

Previous studies [23] adopted a simplified empirical expression, a power-law form (Eq. (4)). This form was motivated by the assumption of a linear relationship between angles ($\Omega' \propto \Omega$), indicating that a theoretical exponent $n = 2$, and their experimental results yielded $n = 2.1$, seemingly confirming this assumption. To further investigate the validity of the simplified power-law approximation, we directly compared our analytical model (Eq. (7), derived from the geometric relation between $\Omega$ and $\Omega'$) to the power-law fitting results under various pillar geometries (Fig. 3(c)). The aspect ratio $\lambda$ is defined as the ratio of the pillar height to diameter ($\lambda = h/d$). For moderate aspect ratios ($\lambda = 1.5$, corresponding to conditions similar to previous studies [22,23]), the power-law approximation closely matches the analytical prediction, with a small error (the root mean square error (RMSE) $= 1.25°$, the coefficient of determination ($R^2$) $= 0.996$). In contrast, under our experimental conditions with a higher aspect ratio ($\lambda = 5.0$), the deviation between the analytical model and the power-law fit was slightly larger ($\text{RMSE} = 4.66°, R^2 = 0.950$), indicating that the power-law approximation became less reliable at high aspect ratios. Thus, our analysis reveals that Eq. (4) remains effective for fitting the experimental data within typical aspect ratio ranges, as in previous studies [23] and this work. However, applying the same power-law approximation for higher aspect ratio pillars may result in increasing deviations from the actual behavior.

To verify the reliability of our analytical model, we applied it to fit the experimental data spanning various pillar spacings and anisotropic array orientations. As shown in Fig. 3(d), which includes the experimental data shown in Fig. 2(c) along with the literature data, the model consistently fits different neighbor types and experimental conditions. Notably, our data in this ⊥ direction shows nearly perfect agreement with the model, indicating that the bridging behavior in this direction is well governed by the geometric constraints formalized in our theory. In contrast, when applying our analytical model to the prior work's dataset [23], we observed excellent agreement for their second NN (2NN) bridging data, whereas moderate deviations appeared in their first NN (1NN) data. This slight discrepancy may be attributed in part to the differences in the definition of bridging in each study. In the referenced work, bridging was counted as a binary outcome—whether a CNT bridged a given neighbor pair or not—regardless of the number of bridging CNTs. In contrast, the overall consistency and accuracy of our analytical model across other datasets, including our measurements, supports its general validity and robustness. It is worth noting that although the analytical bridging model assumes cylindrical pillars for simplicity, the actual structures used in the experiments are square in cross-section. This approximation remains valid because the interpillar distances are much larger than the difference between the square and circular cross-sectional dimensions, resulting in only minor variations in the angular accessibility. Moreover, unlike the square pillars, the cylinder model is rotationally invariant, which makes it more general and applicable regardless of the array orientation. The consistent agreement with the experimental data further supports the validity of the geometric simplification.

To explain the directional dependence observed in the CNT bridging, a vibrational growth model was previously proposed [22] as an extension to the isotropic model. Although the isotropic model provides excellent fits across all datasets, there are inconsistencies when comparing the fitted values of the proportionality constant $K$ within the same sample group. For example, in the // group, $K$ values differ significantly between //NN and ∠NN cases; similarly, the difference between ⊥NN and

∠NN is also significant in the ⊥ group. Since these data were obtained under identical experimental conditions, such differences cannot be attributed to variations in the CNT density. This indicates that the isotropic model alone is insufficient to explain the directional dependence of the bridging behavior. To address this issue, previous studies proposed the CNT vibration interpretation based on the isotropic model. Although their analytical expression remained the same, the authors attributed the variation in $K$ to differences in the CNT vibrational mechanism. Based on the geometric considerations (Fig. 3(b)), the ratio of $K_1$ to $K_2$ can be expressed as follows:

$$\frac{\theta + \Omega_1}{90° - \Omega_1 - \theta} = \frac{\theta_1}{\theta_2} = \frac{B_1(L_1)}{B_2(L_2)} = \frac{K_1 \Omega_1 \Omega_1'}{K_2 \Omega_2 \Omega_2'}, \quad (9)$$

where $K_1$ and $K_2$ represent the fitted proportionality constants for the CNT bridging along the ⊥ and ∠ directions, $\theta$ denotes the effective vibrational angle range of the CNT during growth, $\Omega_1$ and $\Omega_2$ represent the transverse angles corresponding to ⊥NN and ∠NN connections, respectively, $\Omega_1'$ and $\Omega_2'$ represent the longitudinal angle, $\theta_1$ and $\theta_2$ represent the accessible vibrational angular ranges for ⊥NN and ∠NN connections, respectively. As shown in Fig. 3(b), each CNT oscillates within an angular range of $\theta$ and can bridge to neighboring pillars if its trajectory intersects the accessible angular range. In the ⊥ direction, any CNT that swings within the combined angle $\theta + \Omega_1$ will eventually land on a ⊥NN pillar. In contrast, to bridge diagonally, the CNT must avoid the ⊥NN pillar entirely and remain within the narrower diagonal angle defined by $90° - \Omega_1 - \theta$. Assuming a symmetric geometry where $\Omega_1 \approx \Omega_2$, $\Omega_1' \approx \Omega_2'$ and that $\Omega_1$ remains small ($< 10°$), since the pillar diameter is much smaller than the interpillar spacing, the expression simplifies to:

$$\frac{K_1}{K_2} \approx \frac{\theta}{90° - \theta}. \quad (10)$$

In our experiments, the estimated $K$ ratio for ⊥NN is approximately 3, corresponding to a vibration angle of $\theta \approx 68°$, which closely matches the value derived from prior data [23]. Since the ⊥NN connection is aligned ⊥ to the gas flow direction; thus, it is not affected by the flow-directed elongation and can be regarded as predominantly governed by the vibrational mechanism. Moreover, previous studies have reported CNT oscillation angles of around 30°, giving a total swing range of approximately 60°, which is consistent with the 68° estimate, further supporting the validity of the vibrational growth model. In contrast, the //NN connection is aligned horizontally to the gas flow direction and is more susceptible to bridging by the gas-flow-induced force. Therefore, the corresponding angle estimated from the $K$ ratio ($\theta \approx 75°$) should not be interpreted as a pure vibrational amplitude. Instead, it likely reflects a combined contribution from both the vibrational and kite-growth mechanisms.

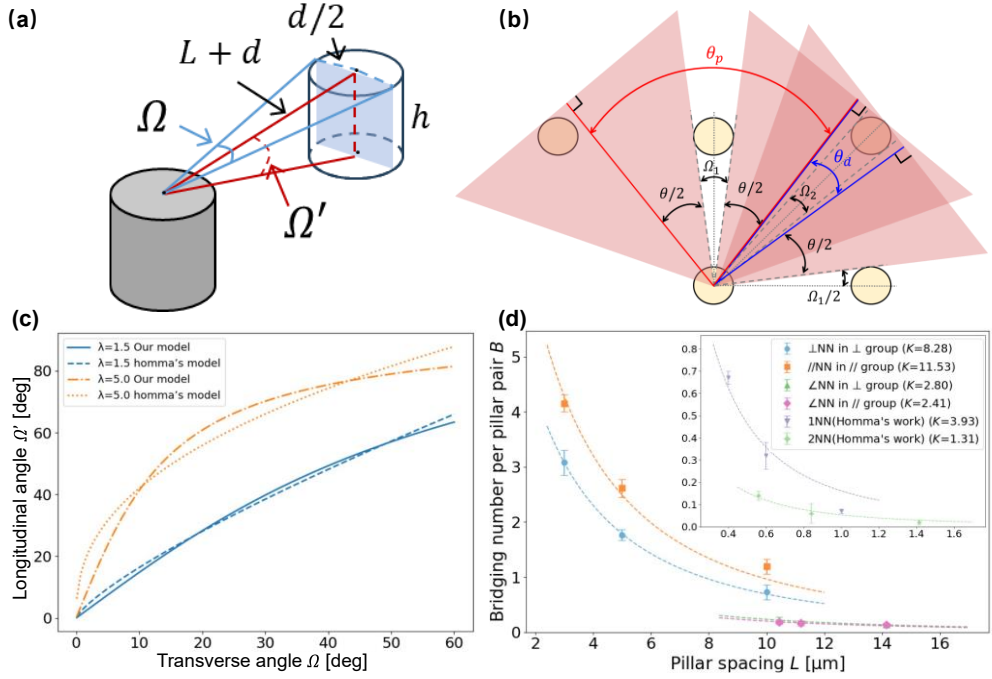

**Figure 3 (a) Schematic illustration of the geometric constraints defining the transverse angle $\Omega$ and the longitudinal angle $\Omega'$ between the nanopillar pairs. (b) Schematic illustration of the geometric model used to derive the ratio of bridging constants $K_1$ to $K_2$. The shaded isosceles triangles represent the vibrational angular range $\theta$ accessible to a CNT growing from the center pillar. $\Omega_1$ and $\Omega_2$ represent the transverse angles corresponding to ⊥NN and ∠NN connections, respectively. $\theta_1$ and $\theta_2$ represent the accessible vibrational angular ranges for ⊥NN and ∠NN connections, respectively. (c) Comparison between the analytical expression of $\Omega'(\Omega)$ and a power-law approximation $\Omega' = k\Omega^n$ under two pillar aspect ratios ($\lambda$ = 1.5 and 5.0). (d) Average bridging number per pillar pair $B$ versus pillar spacing $L$ fitted using the analytical model.**

3.2 Gas flow-induced deflection and vibrations

To further elucidate the influence of the gas flow on the CNT growth behavior, steady-state fluid dynamics simulations based on the FEM were conducted. Fig. 4 shows the simulated velocity fields around the nanopillar arrays with three different configurations: 3 × 10, 5 × 10, and 10 × 10. As shown in the figure, each row corresponds to a horizontal cross-sectional slice at a different vertical height: near the substrate (~4 μm above the base), at the pillar top (8 μm), and above the pillar (~12 μm). The gas flows horizontally from left to right, and the color scale from low (blue) to high (red) represents the velocity magnitude.

Although the velocity in the upper layer above the pillars remains largely unaffected across all configurations, the flow behavior at the pillar tops and near the substrate is more sensitive to the interpillar spacing. These two regions are essential because CNT bridging predominantly occurs near the pillar top and the substrate. At a 3 μm spacing, the dense pillar array significantly impeded the gas flow, thereby reducing the flow velocity between the pillars at the height of the pillar tops to approximately one-half of the inlet velocity and making the flow near the substrate nearly stagnant. When the spacing increased to 5 μm, the obstruction effect diminished. The flow velocity between the pillars at the height of the pillar tops noticeably increases, and small regions even reach the inlet velocity. At the widest spacing of 10 μm, confinement is essentially eliminated because the increased gap between the pillars allows the gas flow to pass through with minimal obstruction. As a result, the flow between the pillars, both near the tops and close to the substrate, achieves an inlet velocity. These results indicate that as the interpillar spacing increases, the influence of the gas flow becomes more pronounced in regions where CNT bridging occurs. This raises a critical question: how does this evolving flow environment impact CNT growth and bridging mechanisms?

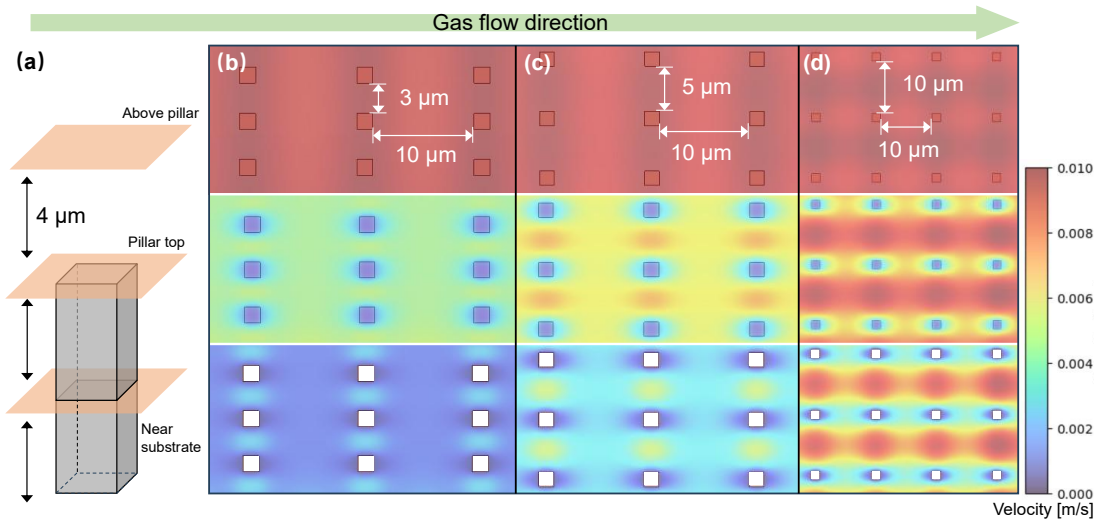

**Figure 4 (a) Schematic illustration of a square nanopillar array and the positions of three horizontal cross-sectional slices: above the pillar (12 μm from substrate), at the pillar top (8 μm), and near the substrate (4 μm). (b–d) Simulated velocity magnitude distributions under horizontal gas flow (left to right) for pillar spacings of 3, 5, and 10 μm, respectively. Each row corresponds to one of the three slice heights defined in (a). Color indicates the magnitude of the flow velocity in arbitrary units, with warmer colors representing higher velocities. Note that in the bottom row (near the substrate), the square regions correspond to the cross-sections of the nanopillars. No fluid flow occurs within these structures, so no velocity is defined inside them. The apparent coloration inside the pillars is due to the material rendering in the simulation**

To quantify the effect of gas flow on individual CNT growth, it is critical to recognize that the CNT diameter is considerably smaller than the mean free path of the gas molecules under typical CVD conditions. Thus, the flow regime corresponds to the free molecular flow characterized by a Knudsen number $K_n = \Lambda/d \gg 1$, where $\Lambda$ denotes the gas mean free path, and $d$ denotes the CNT diameter, rendering conventional continuum-based drag models inapplicable. Accordingly, the drag force $F$ exerted on a CNT can be estimated using an empirical expression suitable for the free molecular flow regime [33]:

$$F = \sqrt{2\pi m_g kT} \left[\varphi + \left(2 - \frac{6-\pi}{4}\varphi\right) \sin^2 \alpha\right] NRL_c V. \tag{11}$$

Here, $m_g$ denotes the mass of a gas molecule, $k$ denotes Boltzmann's constant, $T$ denotes the absolute temperature, $\varphi$ denotes the momentum accommodation coefficient (typically 0.9), $\alpha$ denotes the angle between the gas flow direction and the CNT axis (set to 90°), $N$ denotes the number density of the gas molecules, $R$ denotes the CNT radius, $L_c$ denotes the CNT length, and $V$ denotes the relative gas flow velocity.

The corresponding lateral deflection of the CNT under the influence of this drag force can be modeled by treating the CNT as a cantilever beam subjected to a uniform load. The deflection $\delta$ at the free end is calculated as follows [34]:

$$\delta = \frac{qL_c^4}{8YI}. \tag{12}$$

Here, $q$ denotes the distributed load per unit length given by $F = qL_c$, $Y$ denotes the Young's modulus of the CNT, and $I$ denotes the second moment of the area. Under the flow conditions used in this study, analytical calculations based on Eq. (12) suggest that CNTs with lengths smaller than approximately 1 μm exhibit minor deflection, typically below 1 nm, as summarized in Table S3. In principle, the flexibility of the catalyst particle at the CNT root can influence the amplitude of the CNT tip deflection during growth. However, explicitly incorporating the catalyst deformation into the analytical model would significantly complicate the derivation. To address this issue, FEM simulations were conducted to assess the potential impact of the catalyst flexibility. First, without considering catalyst softening, the FEM results (summarized in Table S3) showed that the tip deflection remained small, with values around 0.059 nm for a CNT with a length of 400 nm. Subsequently, a 5-nm hemispherical iron particle with reduced mechanical stiffness was introduced at the CNT root, where the elastic moduli were reduced to one-tenth of their expected high-temperature values to approximately account for the thermal softening effects. This modification slightly increased the tip deflection from 0.059 to 0.079 nm for a CNT of the same length. However, the overall influence remained limited. Therefore, in the subsequent analyses, the catalyst flexibility was reasonably neglected to simplify the modeling without a significant loss of accuracy. The gas-flow-induced deflection as a function of the CNT length is plotted as the blue curve in Fig. 5(a).

In contrast, vibrations were identified as the main factors that modulated the growth of shorter CNTs. Although previous studies have primarily attributed such vibrations to mechanical vibrations from the CVD system [22,25], it is essential to recognize that thermal excitation may also contribute significantly. In a previous study [22], the vibrational amplitude, defined as the lateral displacement range of the CNT tips during growth, was first calculated using a cantilever beam model, which was found to be insufficient to explain the observed high proportion of NN bridging. To address this issue, the authors proposed that the catalyst particles become molten during growth, which could enhance the large-amplitude motion driven by external mechanical vibrations. This view is further supported by subsequent experimental studies showing that catalyst particles can indeed become molten during growth due to the formation of carbon–metal eutectic compounds in the CVD environment [35]. However, the same softening at the CNT root would also reduce the effective stiffness of the system [36], thereby increasing the vibrational amplitude under thermal excitation, according to the equipartition theorem [37]. To evaluate both possibilities, we derived and compared the analytical expressions for the vibration amplitude arising from the thermal and external mechanical vibrations. The amplitude $\sigma$ of thermal vibrations is given by [38]

$$\sigma = 0.9212 \left( \frac{kT}{YWG(W^2 + G^2)} \right)^{0.5} L_c^{1.5}, \tag{13}$$

where $W$ and $G$ represent the effective cross-sectional dimensions. The red curve in Fig. 5(a) shows the thermal vibration amplitude estimated from the analytical calculations based on Eq. (13).

The amplitude of the mechanical vibration can be estimated by modeling the CNT as a cantilever beam under low-frequency forced vibration conditions. The mechanical vibration originating from the rotary pump is assumed to be directly transmitted to the CNT without attenuation. In other words, the acceleration acting on the CNT is equal to the pump-induced acceleration:

$$a_0 = A_p \cdot (2\pi f)^2, \tag{14}$$

where $A_p$ and $f$ denote the amplitude and frequency, respectively, of the pump vibration. We adopted relatively high values for these parameters—$A_p = 50$ μm and $f = 100$ Hz—which are within the upper range of typical mechanical vibration in CVD systems. Under the low-frequency limit, the vibration amplitude $A$ of the CNT refers to the maximum lateral displacement of its free end during growth and can be approximated as follows [39]:

$$A \approx \frac{F_0}{m_{eff}\omega_0^2} = \frac{F_0}{0.236m\omega_0^2} = \frac{a_0}{0.236\omega_0^2}, \quad (15)$$

where $F_0$ denotes the external force on the CNT, the coefficient 0.236 represents the normalized effective mass of the CNT in its first flexural mode, and $\omega_0$ denotes the natural angular frequency of the first flexural mode given by [37]:

$$\omega_0 = \frac{\beta_0^2}{2L_c^2}\sqrt{\frac{Y(W^2 + G^2)}{2\rho}}. \quad (16)$$

Here, $\rho$ denotes the density of the CNT, and $\beta_0 \approx 1.875$ represents the mode constant of the first flexural mode. By substituting Eqs. (14) and (16) into Eq. (15), the vibration amplitude induced by the mechanical vibration from the CVD system is given by

$$A = \left(\frac{32\pi^2\rho f^2 A_p}{0.236\beta_0^4 Y(W^2 + G^2)}\right)L_c^4. \quad (17)$$

The results in Fig. S4 show that even under conservative assumptions, the mechanical vibrations remain significantly smaller than those induced by the thermal vibrations under typical growth conditions. Therefore, we consider the thermal vibration to be the dominant vibrational mechanism that influences the CNT bridging, whereas the contribution of the externally induced mechanical vibrations can be reasonably regarded as negligible.

Based on this understanding, we further compared the magnitudes of the gas-flow-induced deflection and thermal vibration amplitude as functions of the CNT length to identify a distinct crossover behavior. This crossover is inevitable, as the thermal vibration amplitude scales with the CNT length as $L_c^{1.5}$, whereas the gas-flow-induced deflection grows much faster, scaling as $L_c^4$. Fig. 5(a) shows this relationship under the growth conditions used in our experiments. For CNT lengths shorter than approximately 4 μm, the thermal vibration dominates the tip motion, and the bridging behavior is primarily governed by the intrinsic vibrational dynamics around the catalyst. This regime favors short-range 1NN connections and is largely insensitive to gas flow. However, for CNTs longer than ~4 μm, flow forces induce greater deflection, and kite growth becomes the dominant mechanism, enabling longer-range, flow-aligned bridging paths.

Importantly, the crossover point between these two regimes is not fixed; it varies with experimental conditions like temperature and gas flow rate. Fig. 5(b) compares the crossover length under different growth conditions used in previous studies [23]. Due to differences in the experimental conditions—specifically, a higher gas flow velocity, lower system pressure, and lower temperature—the crossover point in their system occurs at a longer CNT length than the actual length grown in their experiments. This explains their conclusion that the gas flow plays little to no role in CNT bridging. Under these conditions, the CNTs remained too short to be affected by the flow-induced forces. In contrast, CNTs grow sufficiently long in our system to exceed the crossover length, making the gas flow a decisive factor in the growth dynamics. In addition to the previous study [23], another previous study [16] also provided important evidence supporting the crossover model proposed in this work.

Fig. 5(c) shows the crossover length under the growth conditions reported in [16]. Under these conditions, the crossover point appeared at a CNT length between the lengths of the short CNTs forming the network and the longer CNTs observed bridging multiple pillars. For short CNT networks, this is consistent with their SEM observations, where CNTs bridging the // and ⊥ directions appear in similar numbers, indicating the absence of strong directional alignment with the gas flow. Additionally, when a CNT bridges from the top of one pillar to a neighboring pillar, the attachment points are distributed across the top, middle, and bottom regions of the receiving pillar. This spatial distribution suggests that the CNT tips oscillate due to the thermal vibrations during growth rather than the flow-induced force. In contrast, for isolated ultralong CNTs exceeding 100 μm, flow-induced forces were dominant. In these cases, the CNT midsections were mostly found lying near the tops of the adjacent pillars, which is consistent with the kite-growth mechanism, in which the CNTs are lifted by the gas flow and fall onto the neighboring structures from above. These observations provide strong evidence that both vibrational and kite-growth modes are present, depending on the CNT length, which aligns well with the crossover model proposed in this work.

Fig. 5(d) shows how the crossover point between the thermal-vibration-dominated and gas-flow-dominated regimes varies with the gas-flow velocity. For comparison, the effects of temperature and pressure on the crossover length were also evaluated (Fig. S5). Although these parameters influence the crossover point, their impact is notably less significant than the gas flow velocity. Under increasing flow rates, the aerodynamic drag exerted on the CNTs significantly increased, accelerating the onset of the flow-induced deflection relative to the thermally induced vibration amplitude. As a result, the length at which the gas flow begins to dominate over the thermal

motion—the crossover length—shifts progressively to lower values. For example, at a low flow velocity of 0.002 m/s, crossover occurs at approximately 6 μm, indicating that only relatively long CNTs are subject to flow-dominated behavior. As the flow velocity increased to 0.01 m/s, the crossover point decreased to approximately 4 μm. Further increasing the flow velocity to 0.05 m/s or a faster gas flow velocity shifts the transition point down to the 2–3 μm range. This means that under a strong gas flow, even short CNTs are susceptible to aerodynamic deflection, enabling the kite-growth mechanism to engage earlier during growth. These results highlight the tunability of the CNT growth regime via the gas flow control. Therefore, flow velocity is not just a passive growth parameter—it actively governs the nature and scale of the CNT network formation. In the absence of gas flow during growth, the crossover length can be considered effectively infinite, indicating that the entire CNT growth process is governed by a vibrational mechanism. This limiting case was observed in previous studies and agreed well with the vibrational growth model [23], as shown in the fitting results in Fig. S6.

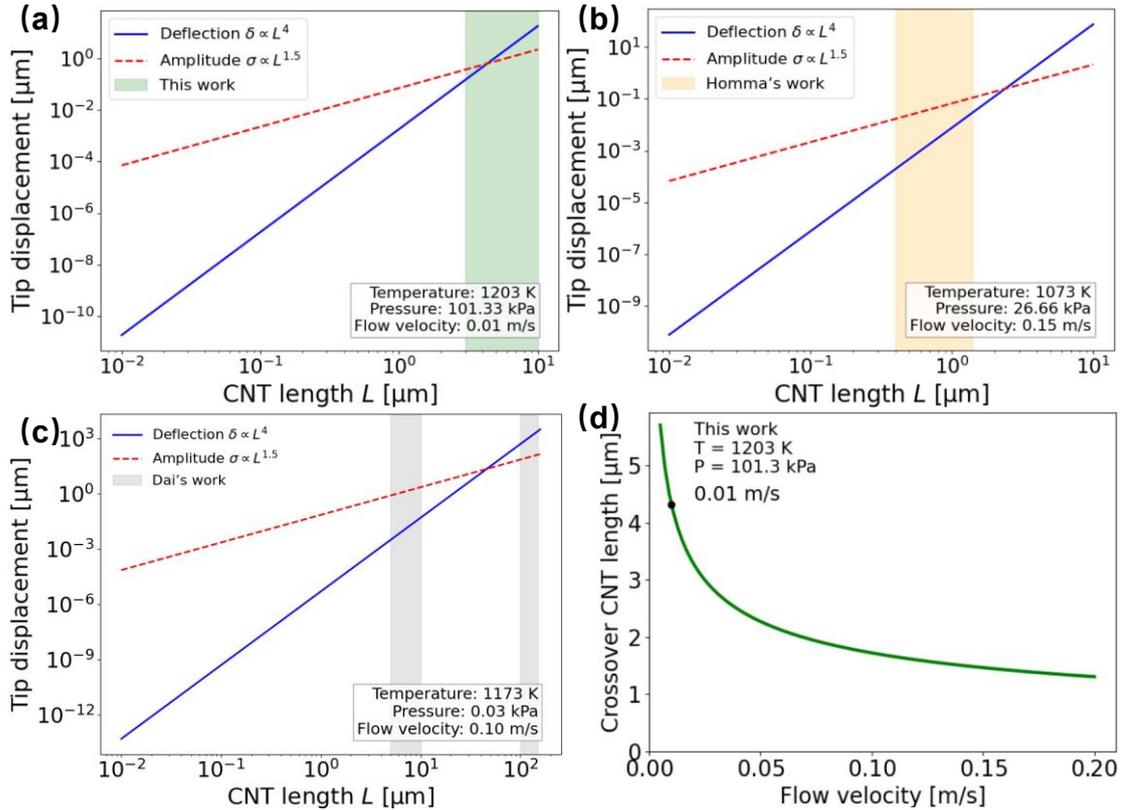

**Figure 5 (a, b, c) Log–log plot comparing the gas-flow-induced deflection (blue line) and thermal vibration amplitude (red dashed line) as a function of the CNT length under different experimental conditions: (a) this work, (b) a previous study [23], and (c) another previous study [16]. The shaded regions represent the typical CNT length range under each condition. (d) The crossover length shifts with increasing gas flow velocity. The curves were generated**

**from the model calculations under the experimental conditions used in this work.**

3.3 Statistical analysis of the CNT bridging types

To comprehensively characterize the structural properties of the CNT networks formed under various growth conditions, we also classified bridging CNTs that do not fall into the previously defined //NN, ⊥NN, or ∠NN categories as "other." These "other" CNTs refer to those that still form suspended bridges but fall outside the defined NN types (//NN, ⊥NN, and ∠NN). They may span more than two pillars, follow highly bent or twisted paths between pillars, or even form entangled structures with neighboring CNTs, making them distinctly different from straight NN bridges. Although these CNTs are relatively rare, their presence becomes more noticeable under kite growth conditions, which facilitates longer CNT extensions. These bridges contribute to increased network complexity and may enhance signal propagation diversity, potentially contributing to applications, such as physical reservoir computing and neuromorphic circuits [40–44].

Fig. 6 categorizes the CNT bridges into three types. The first category is the //NN bridges, which align with the gas flow direction and are influenced by kite growth and vibrations. The second is the ⊥NN bridges, which are difficult to align due to gas-flow-induced forces, primarily formed due to vibrations. The third category is the ∠NN and "other" bridges, which generally span greater distances, making vibrations negligible, although they are influenced by the gas flow (Fig. 5(a)). A notable indication of kite growth can be observed when comparing the 3 × 10 and 10 × 3 layouts. Despite having the same short pillar spacing (3 μm), the 3 × 10 layout shows a higher density of //NN bridges compared to the ⊥NN bridges in the 10 × 3 layout. Given that the vibrations remained consistent in both instances, this difference can be attributed to the directional contributions of the kite growth, allowing the CNTs to extend longer and align better with the gas flow. The numerical values related to Fig. 6 are summarized in Table S4.

As shown in the 3 × 10 vs. 10 × 3 and 5 × 10 vs. 10 × 5 layouts, the actual density of ∠NN and "other" bridges remains nearly constant regardless of the flow direction. This suggests that their formation is not influenced by the direction of the gas flow. These bridges tend to be longer and can be influenced by flow-induced effects; however, they exhibit an angular deviation from the flow direction. In such cases, the gas flow may even create disturbances that lead to deflected or bent connections, contributing to the ∠NN and "other" types. Thus, their actual density is primarily influenced by the arrangement of the pillars rather than the flow alignment. However, their percentage relative to the total number of CNT bridges does change with variations in the gas flow direction. The 3 × 10 layout shows an increase in //NN bridges, which reduced the proportion of ∠NN and "other" types. Although the actual density in the 3 × 10 layout exceeded that of the 10 × 10 layout, the proportion decreased to below 5%, compared to almost 10% in the 10 × 10 layout. This indicates that both the gas flow direction and the arrangement of pillars can be adjusted to control the density and relative proportions of ∠NN and "other" types, offering a level of structural tunability during network formation.

Notably, the bridge-type distribution reported in the previous study [22], which used iron catalysts, was close to our 5 × 10 result. Specifically, their CNT networks contained 86% NN bridges, 9% second-NN bridges, and 5% classified as "other." These values closely match our 5 × 10 data: 87% //NN + ⊥NN (NN), 8% ∠NN (second-NN), and 5% "other." Although the numbers are similar, the growth mechanisms differ. In their study, the pillar arrangement was fixed, and they only experienced vibrational growth. In contrast, our findings indicate that by controlling the pillar arrangement and gas flow, we can consistently achieve bridge-type distributions comparable to those associated only with vibrational growth. These findings highlight how the pillar arrangement and gas flow influence the CNT bridging behavior. Kite growth enhances directional alignment and enables long-range connectivity, whereas the pillar arrangement influences vibrational bridging for short-range CNT connections. Additionally, the growth conditions employed in this study were originally optimized to promote kite growth and facilitate the formation of long, suspended CNTs. However, in other experiments with varying gas flow rates under identical growth conditions, we found that the CNT bridging behavior and connection-type distribution could be altered. These findings suggest that optimal growth conditions for maximizing the CNT length do not necessarily yield the most structurally balanced or morphologically diverse networks. Therefore, it is crucial to refine the growth conditions not only for producing long CNTs but also for promoting controllable bridging behavior tailored to the desired network structure.

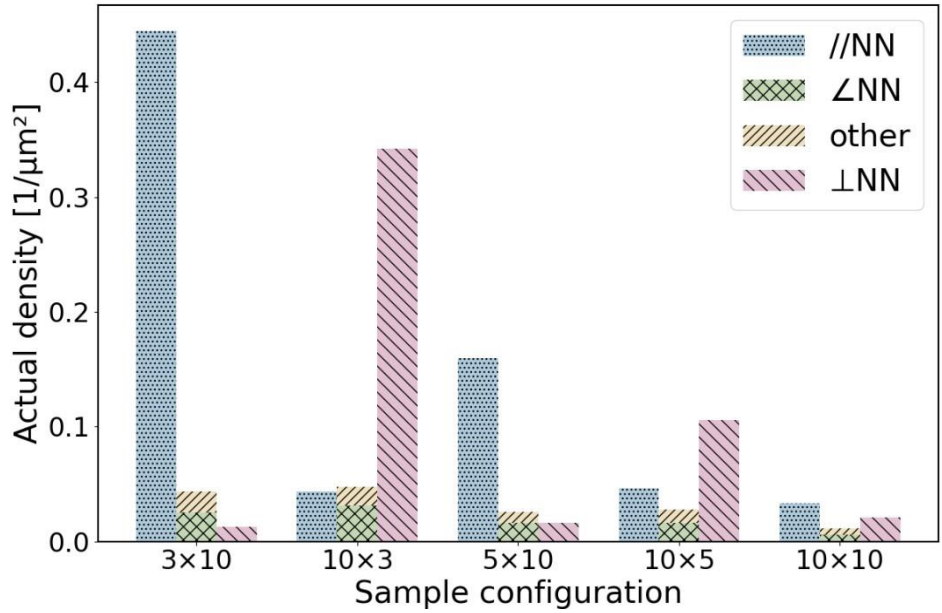

**Figure 6 Actual density distributions of the CNT bridges for different sample configurations. Stacked bar chart depicting the contributions of //NN, ⊥NN, ∠NN, and "other" CNT bridges to the total network density. The //NN bridges are aligned in the direction of the gas flow and are affected by kite growth and vibrations. The ⊥NN bridges, which are difficult to align with gas-flow-induced forces, mainly form because of vibrations. In contrast, ∠NN and "other" bridges tend to cover larger distances, rendering vibration insignificant, and they are still influenced by gas flow.**

## 4. Conclusion

In this study, we revisited pillar-assisted CNT growth and developed a generalized analytical model to describe the vibrational bridging of CNTs between nanopillars. By incorporating geometric constraints, the proposed model extends the previous vibrational model and quantitatively captures the relationship between the pillar spacing and bridging number. Introducing the kite growth mechanism allowed us to reveal a crossover behavior between the two competing growth modes. Although the vibrational model governs short CNTs, longer CNTs are increasingly influenced by the gas-flow-induced deflection. The position of this crossover point depends on the CNT length and growth conditions and can be systematically tuned through the gas flow rate and other conditions. Statistical analysis of the CNT bridging further demonstrates that the structural composition of the resulting network can be controlled by adjusting the gas flow direction and pillar arrangement. These findings provide a unified understanding of CNT network formation under kite-guided, pillar-assisted growth and offer practical design guidelines for nanostructured interconnects and wiring applications. Moreover, these structurally tunable networks show potential as a platform for future physical reservoir computing systems [40–44], where complex connectivity and material nonlinearity offer key functional advantages.

## Acknowledgments

The authors would like to thank Dr. T. Sakata of the Research Center for Ultra-High Voltage Electron Microscopy, Osaka University, for assistance in the SEM observation. A part of this work was financially supported by JSPS, KAKENHI (Grant Numbers JP15H05867, JP17H02745, and JP24K01297) and JST, CREST (Grant Number JPMJCR20B5). A part of this work was conducted at the Advanced Research Infrastructure for Materials and Nanotechnology Open Facilities, Osaka University (Grant Number JPMXP1225OS1006).

junction resistance on spatiotemporal dynamics and reservoir computing performance arising from an SWNT/POM 3D network formed via a scaffold template technique, Nanoscale 15 (2023) 8169–8180. https://doi.org/10.1039/D2NR04619A.

Supporting information

# Crossover between kite growth and vibrational bridging in pillar-assisted controlled formation of carbon nanotube networks

*Yuanjia Liu [a], Taiki Inoue [a], Yoshihiro Kobayashi [a]*

a Department of Applied Physics, The University of Osaka,

Suita, Osaka 565-0871, Japan

# S1. Geometric model derivation

To analytically estimate the average number of CNT bridges $B$, we consider the transverse angle $\Omega$ and the longitudinal angle $\Omega'$ that describe the accessible angular space between neighboring pillars. Based on the geometrical relationships illustrated in Fig. S1, the bridging number is expressed as:

$$B = k\Omega \cdot k'\Omega' = K\Omega\Omega', \tag{S1}$$

where $K$ is a proportionality constant determined by the CNT density and growth conditions.

1. Derivation of $\Omega$

From the geometric configuration, the transverse angle $\Omega$ in the horizontal plane is given by:

$$\Omega = 2\arcsin\left(\frac{d}{2(L+d)}\right). \tag{S2}$$

This relationship follows from:

$$\sin(\Omega/2) = \frac{d}{2(L+d)}, \tag{S3}$$

$$\tan(\Omega/2) = \frac{d}{2y}, \tag{S4}$$

$$\cos(\Omega/2) = \frac{z}{y}. \tag{S5}$$

2. Derivation of $\Omega'$

The the longitudinal angle $\Omega'$, which quantifies the vertical accessibility range for bridging, is related to the pillar height $h$ and lateral distance $z$ by:

$$\tan(\Omega') = \frac{h}{z}. \tag{S6}$$

By substituting Eq. (S4) and (S5) into Eq. (S6), and expressing all quantities in terms of $\Omega$, we obtain:

$$\Omega' = \arctan\left(\frac{2h\tan(\Omega/2)}{d\cos(\Omega/2)}\right) = \arctan\left(\frac{2h\sin(\Omega/2)}{d(1-\sin^2(\Omega/2))}\right). \tag{S7}$$

Alternatively, $\Omega'$ can also be expressed directly as:

$$\Omega' = \arctan\left(\frac{h(L+d)}{(L+d)^2 - (d/2)^2}\right). \tag{S8}$$

3. Final Expression for Bridging Number $B$

By substituting Eq. S(2) and (8) into (1), we arrive at the analytical form:

$$B = 2K \cdot \arcsin\left(\frac{d}{2(L+d)}\right) \cdot \arctan\left(\frac{h(L+d)}{(L+d)^2 - (d/2)^2}\right), \qquad (S9)$$

This equation shows that the bridging number $B$ depends on the geometric parameters of the system: pillar spacing $L$, pillar diameter $d$, and pillar height $h$. Notably, if the aspect ratio $h/d$ is fixed, the expression can be fully described as a function of $\Omega$, simplifying the angular modeling framework.

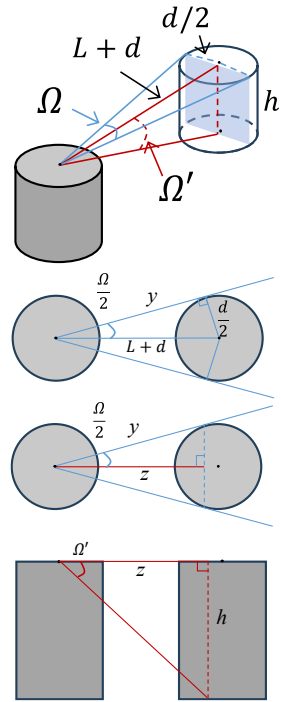

**Figure S2 Geometric model used to derive angular accessibility between CNTs and neighboring pillars.**

## S2. Supplementary figures

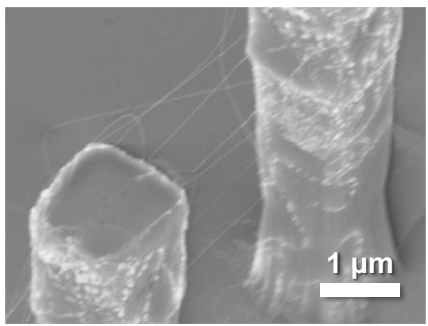

**Figure S3** SEM image of CNT bridges between two nanopillars.

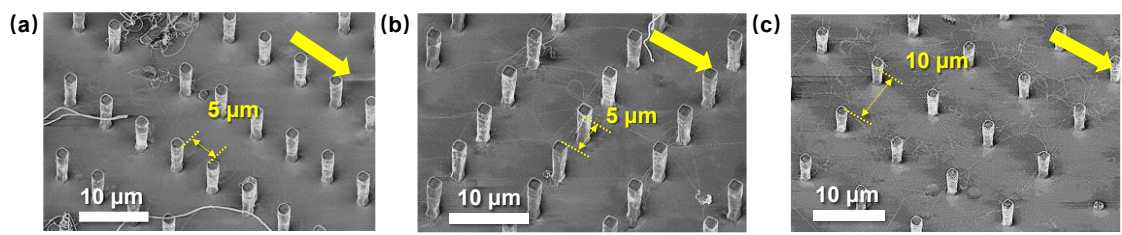

**Figure S4** SEM images of CNT networks grown on substrates. (a) Sample with 5 μm spacing in the // direction and 10 μm spacing in the ⊥ direction; (b) Sample with 10 μm spacing in the // direction and 5 μm spacing in the ⊥ direction. (c) Sample with 10 μm spacing in the // direction and 10 μm spacing in the ⊥ direction. Yellow arrows indicate the direction of gas flow during growth.

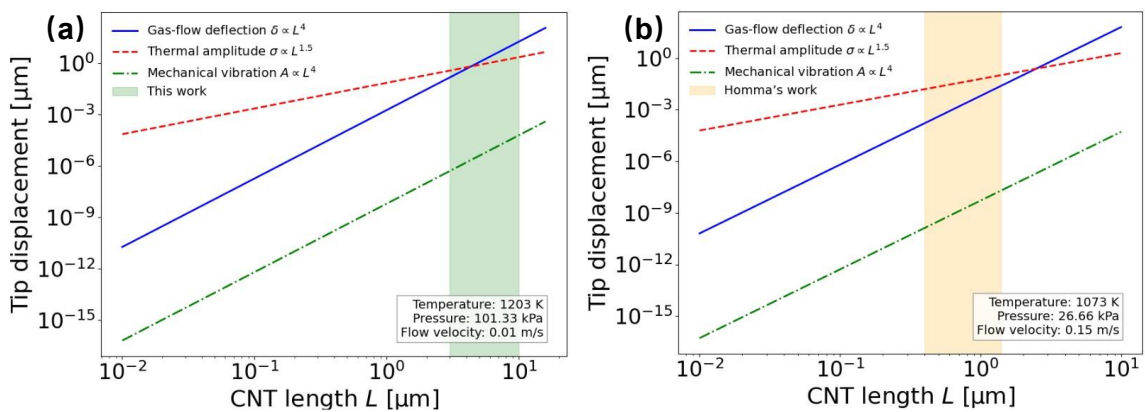

**Figure S5** (a) Log-log plot comparing gas-flow-induced deflection (blue line), thermal vibration amplitude (red dashed line) and vibration amplitude induced by mechanical vibration (green dashed line) as a function of CNT length under different experimental condition: (a) this work, (b) a previous study.

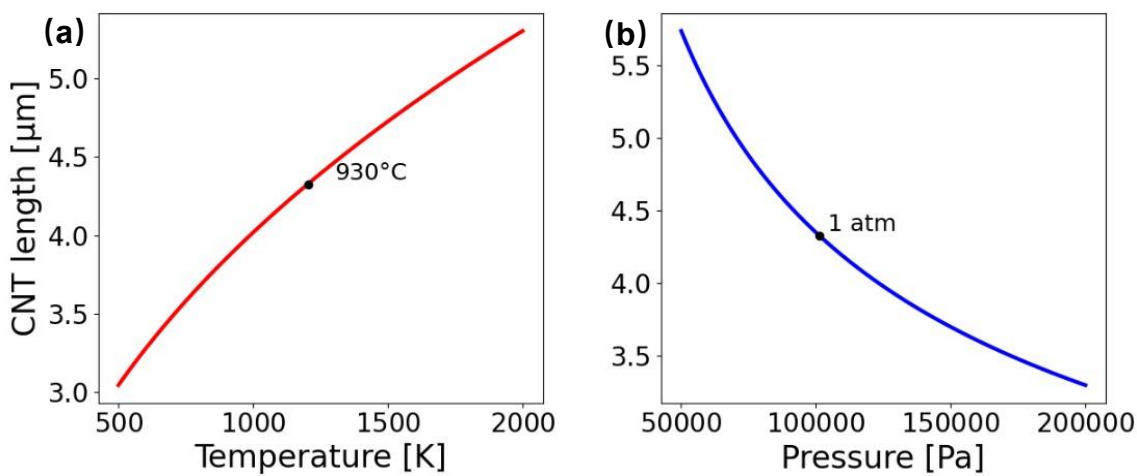

**Figure S6** Dependence of the crossover CNT length on (a) temperature and (b) pressure.

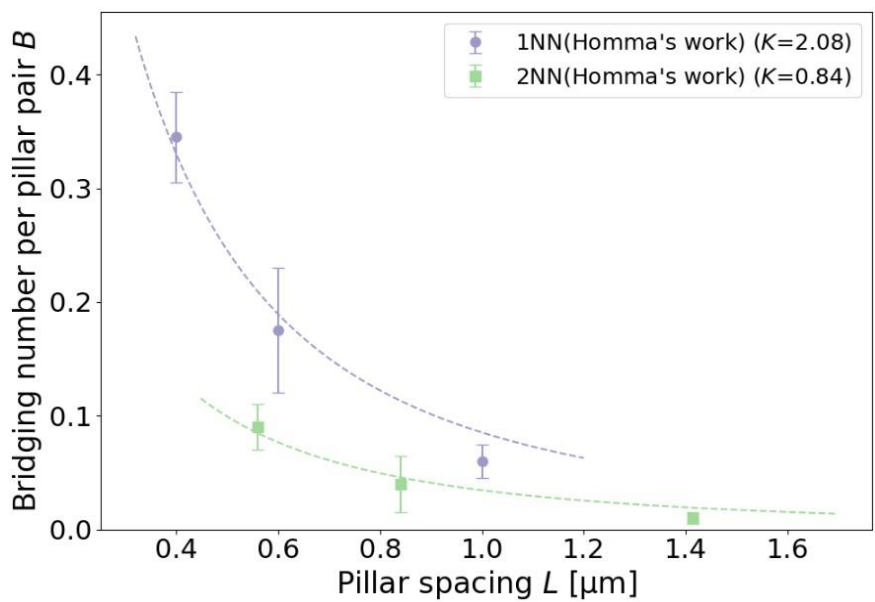

**Figure S7 Average bridging number per pillar pair $B$ versus pillar spacing $L$ fitted using the analytical model under no gas flow condition.**

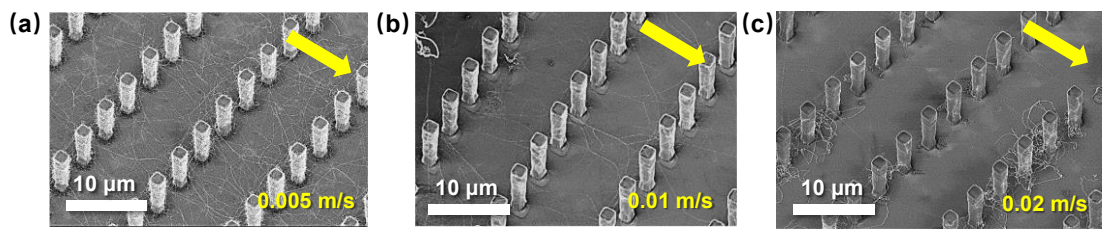

**Figure S8 SEM images of CNT networks grown at 930 °C under different gas flow velocities: (a) 0.005 m/s, (b) 0.01 m/s, and (c) 0.02 m/s.**

# S3. Supplementary Tables

**Table S1 Gas flow simulation**

| Parameter | Value |
|---|---|
| Physics module | Laminar Flow |
| Flow condition | Steady-state, incompressible |
| Gas density (ρ) | 0.404 kg/m³ (Ar) |
| Dynamic viscosity (μ) | $2 \times 10^5$ Pa·s |
| Inlet velocity | Set to match experimental flow rates |
| Boundary conditions | No-slip at nanopillar surfaces, outflow at outlet |
| Mesh strategy | Extra fine tetrahedral mesh near nanopillars; element size ~20 nm |
| Solver | Direct, relative tolerance 1e-5 |

**Table S2 CNT mechanical simulation**

| Parameter | Value |
|---|---|
| Shell Interface (CNT Model) | |
| Shell thickness | 0.34 nm |
| Density | 2000 kg/m³ |
| Axial Young's modulus | $1 \times 10^{12}$ Pa |
| Radial Young's modulus | $5 \times 10^{10}$ Pa |
| Poisson's ratio | 0.16 |
| Shear modulus | $1 \times 10^{10}$ Pa |
| Geometric nonlinearity | Enabled |
| Load | Uniform lateral load |
| Solid Mechanics (Fe Catalyst) | |
| Diameter | 5 nm |
| Density | 6700 kg/m³ |
| Bulk modulus | $1.5 \times 10^{10}$ Pa |
| Shear modulus | $4.5 \times 10^9$ Pa |
| Boundary condition | Base fixed |
| Mesh strategy | Minimum ~0.2 nm |
| Solver | Nonlinear stationary, relative tolerance 1e-5 |

**Table S3 Estimated CNT tip displacement under different driving mechanisms**

| Driving mechanism | CNT length | CNT displacement | |
|---|---|---|---|
| | | Formula (Eqs. (12) and (13)) | FEM simulation |
| Flow disturbance | 400 nm | 0.047 nm | CNT: 0.059 nm |
| | | | CNT+Fe: 0.079 nm |
| | 1 μm | 1.8 nm | – |
| | 4 μm | 470 nm | – |
| Thermal vibration | 400 nm | 18.0 nm | CNT: 16.80 nm |
| | | | CNT+Fe: 40.20 nm |
| | 1 μm | 71.0 nm | – |
| | 4 μm | 568 nm | – |

**Table S4 CNT bridging type proportions and total density across different nanopillar configurations**

| sample | // Spacing (μm) | ⊥ Spacing (μm) | //NN (%) | ⊥NN (%) | ∠NN (%) | Other (%) | Total Density (1/μm²) |
|---|---|---|---|---|---|---|---|
| 3 × 10 | 3 | 10 | 88.8 | 2.6 | 5 | 3.7 | 0.501 |
| 10 × 3 | 10 | 3 | 10 | 79 | 7.2 | 3.8 | 0.433 |
| 5 × 10 | 5 | 10 | 79.3 | 8 | 8 | 4.8 | 0.201 |
| 10 × 5 | 10 | 5 | 25.6 | 58.9 | 8.9 | 6.7 | 0.18 |
| 10 × 10 | 10 | 10 | 50.9 | 32.1 | 9.4 | 7.5 | 0.066 |